# Whistler-mode waves upstream of Saturn

A. H. Sulaiman,[1] D. A. Gurnett,[1] J. S. Halekas,[1] J. N. Yates,[2] W. S. Kurth,[1] M. K. Dougherty[2]

___________

Corresponding author: A.H. Sulaiman, Department of Physics and Astronomy, University of Iowa, Iowa City, Iowa, USA. (ali-sulaiman@uiowa.edu)

[1]Department of Physics and Astronomy, University of Iowa, Iowa City, Iowa, USA.

[2]Space and Atmospheric Physics, Blackett Laboratory, Imperial College London, London, UK.

## Abstract

Whistler-mode waves are generated within and can propagate upstream of collisionless shocks. They are known to play a role in electron thermodynamics/acceleration and, under certain conditions, are markedly observed as wave trains preceding the shock ramp. In this paper, we take advantage of Cassini's presence at ~10 AU to explore the importance of whistler-mode waves in a parameter regime typically characterized by higher Mach number (median of ~14) shocks, as well as a significantly different IMF structure, compared to near Earth. We identify electromagnetic precursors preceding a small subset of bow shock crossings with properties which are consistent with whistler-mode waves. We find these monochromatic, low-frequency, circularly-polarized waves to have a typical frequency range of 0.2 – 0.4 Hz in the spacecraft frame. This is due to the lower ion and electron cyclotron frequencies near Saturn, between which whistler waves can develop. The waves are also observed as predominantly right-handed in the spacecraft frame, the opposite sense to what is typically observed near Earth. This is attributed to the weaker Doppler shift, owing to the large angle between the solar wind velocity and magnetic field vectors at 10 AU. Our results on the low occurrence of whistler waves upstream of Saturn also underpins the predominantly supercritical bow shock of Saturn.

**Introduction**

Whistler-mode waves are low frequency, right-handed, and circularly-polarized electromagnetic waves. They obey a dispersion relation in a two-fluid model which is restricted to the frequency range between the ion and electron cyclotron frequencies, i.e. $f_{ci} \lesssim f \ll f_{ce}$ [*Gurnett*, 1995]. For decades, whistlers have been found in planetary systems deep in the magnetosphere [e.g. *Wu et al.*, 1983] and upstream of the bow shock alike [e.g. *Fairfield*, 1974]. In the latter, they are observed as precursors which grow from the nonlinear steepening of right-handed waves as the length scale shortens ($k$ increases). The direction of disturbances that lead to a whistler-mode transition between the upstream and downstream orientations was found to be consistent with the magnetic field noncoplanarity component within the shock layer [*Thomsen et al.*, 1987].

Whistler precursors, in principle, are associated with low-Mach-number shock waves dominated by dispersion. In contributing to limiting the steepening of a (fast) magnetosonic wave, whistlers carry energy from the shock into the upstream region since they can propagate faster than the magnetohydrodynamic (MHD) waves. The signature for such shocks is manifested as a wave train in the magnetic field preceding the ramp, with a length scale several times greater than the shock thickness. There is, in concert with the dispersion, an associated dissipation understood to be Landau damping from the interaction of electrons and the parallel electric field of the whistler precursor [*Mellott and Greenstadt*, 1984]. *Sundkvist et al.* (2012) experimentally found the whistlers' Poynting flux to be directed upstream, starting at the ramp, thereby revealing the inherent connection of these waves to the dispersive shock structure.

There are two critical Mach numbers associated with quasi-perpendicular collisionless shocks. The first represents the upper limit in solar wind flow speed at which the shock has the capacity to sufficiently dissipate the incident flow by anomalous resistivity and/or wave dispersion [*Marshall*, 1955; *Kennel et al.*, 1985]. Above this, the shock adopts an alternative mechanism, namely ion reflection, to convert the excess ram energy into thermal energy [*Paschmann et al.*, 1981, *Treumann*, 2009]. The second critical Mach number, $M_w$, corresponds to the maximum propagation speed at which dispersive whistlers can stand upstream. This is given by

$$M_w = K \sqrt{\frac{m_p}{m_e}} \cos \theta_{Bn} \qquad (1)$$

where $m_p$ and $m_e$ are the proton and electron masses respectively and $\theta_{Bn}$ is the acute angle the upstream magnetic field makes with the local shock normal (pointing upstream) [*Krasnoselskikh et al.*, 2002]. The constant $K$ depends on the propagation speed in question: ½ for the phase speed, $v_p$, and $\sqrt{27/8}$ for the group speed, $v_g$. There also exists a nonlinear variant of $M_w$ above which a quasistationary whistler wave train is unstable and marks the transition to nonstationary behavior of the shock front [*Krasnoselskikh et al.*, 2002, 2013; *Lobzin et al.*, 2007]. For the purpose of this work, we cannot experimentally make a distinction between the three $M_w$ and shall use the group speed $M_w$.

The first observations of whistler-mode waves upstream of a bow shock took place at Earth [*Fairfield and Feldman*, 1975] and further observations at other planets followed at Mercury [*Fairfield and Behannon*, 1976], Venus [*Orlowski and Russell*, 1991] and Saturn [*Orlowski et al.*, 1992]. *Russell* (2007) conducted a comparative review on whistlers upstream of planetary bow shocks that revealed the dependence of their frequency on heliocentric distance. The author found a monotonic decrease of the (peak) frequency of whistler-mode waves with the decreasing magnetic field strengths upstream of Mercury, Venus, Earth, and Saturn. Moreover, the wave amplitude was shown to decay with distance from the shock (towards upstream) with Landau damping expected to be responsible. Whistlers have also been observed upstream of asteroids [*Gurnett*, 1995], interplanetary shocks [*Wilson et al.*, 2012], and near lunar crustal magnetic sources [*Halekas et al.*, 2006] underlining their ubiquity in collisionless space plasmas.

In this paper, we have studied crossings of Saturn's bow shock between 2007 and 2014 by the Cassini spacecraft to assess the occurrence of whistler waves upstream of quasi-perpendicular shocks. We seek to identify, compare, and explain the differences between their properties in the unique conditions of the solar wind at 10 AU and the more familiar near-Earth space.

**Data and Observations**

We have identified 401 out of the 425 shock crossings as quasi-perpendicular (i.e. $\theta_{Bn} \geq 45°$) and estimated the Alfvén Mach number, $M_A$, for each. The shock normal directions were determined geometrically using a semi-empirical model by *Went et al.* (2010). The reliability of this approach was confirmed using multi-spacecraft timing by *Horbury et al.* (2001) at the terrestrial bow shock. The detailed methods used to determine the shock normal and $M_A$ at Saturn can be found in *Sulaiman et al.* (2016). The high probability of Saturn's bow shock to have a quasi-perpendicular configuration is by virtue of the Parker spiral. The angle the interplanetary magnetic field (IMF) makes with the Sun-planet line increases with heliocentric distance; such that at 10 AU the angle between the upstream IMF and Saturn's (dayside) bow shock is usually large. The data presented in this paper are from Cassini's fluxgate magnetometer [*Dougherty et al.*, 2004] with a sampling rate of up to 8 Hz. The coordinate system is the Cartesian Kronocentric Solar Magnetic (KSM) defined as Saturn-centered with a positive $x$ toward the Sun, $z$ in the plane containing $x$ and Saturn's magnetic axis in the northward sense, and $y$ completing the right-handed system. Evidence for whistler-mode waves was also checked using the search coil of the Radio and Plasma Wave Science (RPWS) instrument [*Gurnett et al.*, 2004].

Figure 1 plots the root-mean-square (RMS) of the components of the magnetic field intervals (~5 mins) upstream, i.e. preceding the ramp and foot and up to the steady IMF, of each quasi-perpendicular shock crossing against $M_A/M_w$. The critical Mach number ($M_A/M_w = 1$) is indicated, separating the subcritical ($M_A < M_w$) from the supercritical ($M_A > M_w$) classes. The fluctuations in the intervals, from which the RMS values were calculated, were band-pass-filtered between the frequency range $f_{ci} < f < f_N$; where $f_{ci}$ and $f_N$ are the ion cyclotron and Nyquist frequencies [*Oka et al.*, 2006]. This frequency range corresponds to that at which whistler waves are expected to be identified in Saturn's foreshock based on the dispersion relation and limited by the sampling rate. Consistent with the theory of the whistler critical Mach number, the RMS exhibits a general trend that falls dramatically near the $M_A/M_w = 1$ boundary. Continuous shock motion in response to the unsteady nature of the solar wind is likely to be the main reason why the sharp drop is not exactly coincident with the critical Mach number. The main errors stem from the sensitivity to $\theta_{Bn}$ which increase rapidly as $\theta_{Bn} \rightarrow 90°$, as can be seen from Equation 1. A single spacecraft means additional errors are induced by shock motion which cannot be minimized, nonetheless a large sample size as such should reveal an underlying general trend among the intrinsic variability. Observationally, this technique is successful in separating shocks into classes of sub- to supercritical regimes with respect to the second (whistler) critical Mach number [e.g. *Oka et al.*, (2006)]. It is clear that the predominant state of Saturn's bow shock is supercritical with respect to both the first and second (whistler) critical Mach numbers [*Sulaiman et al.*, 2015]. This is by virtue of both a higher Mach number in the upstream solar wind and more strongly perpendicular shocks.

Since Cassini is a single spacecraft, we are restricted to determining the properties of whistlers in the spacecraft frame instead of the more convenient plasma rest frame. This leads to a Doppler shift being imposed on the frequencies and polarizations measured. The extent of the Doppler shift's effect on the handedness depends on both $v_p$ and the angle, $\theta_{kv}$, between the wave vector, $k$, and the upstream solar wind vector, $V_{SW}$. Assuming that they are generated at the shock, whistler waves can only be detected in the spacecraft frame when $v_g > V_{SW}$ i.e. the speed of the information can overcome the downstream-travelling solar wind. The polarization is observed as the intrinsic right hand when $v_p > V_{SW}$. Conversely, the polarization is left-handed when $v_p < V_{SW}$ since, while the information is still able to propagate upstream, the phases are swept past the spacecraft and reversed by the more dominant $V_{SW}$. Note that in the dispersion relation, $v_g > v_p$ always.

We have identified 24 crossings with a monochromatic wave train upstream; all in or near the subcritical regime. Figure 2a is an example of a 10-minute magnetic field time series with a 5-second moving-average overlaid. The shock is subcritical with an estimated $M_A/M_w = 0.6$. The upstream wave train (underlined by the dashed line) is background-subtracted in Figure 2b and minimum variance analysis (MVA) is used to transform the coordinate system into the principal axes [*Sonnerup and Cahill*, 1967]. The diagonal variance matrix yields three eigenvalues, $\lambda_{1,2,3}$ whose eigenvectors correspond to the directions of maximum, intermediate, and minimum variances respectively. The ratios of the eigenvalues are such that $\lambda_1/\lambda_2 \sim 1$ indicates a nearly circularly-polarized wave and $\lambda_2/\lambda_3 > 10$ indicates a well-defined propagation direction (note the minimum variance eigenvector corresponds to that direction). Figures 2c and 2d present hodograms along the directions of the minimum and maximum variance eigenvectors respectively. Taking into account the mean magnetic field, the propagation direction with respect to the magnetic field, $\theta_{kB}$ is inferred as $33 \pm 2°$ and with respect to the solar wind vector, $\theta_{kV}$ as $64 \pm 2°$ (the propagation direction is ambiguous in sign, so all angles are referenced to the range of 0-90°). The wave is right-handed polarized with respect to the mean magnetic field with a peak frequency 0.26 Hz in the spacecraft frame. We have identified ~5 % of the total shock crossings to exhibit whistler signatures. The peak frequencies (of maximum power) were distinct in the power spectrum of the magnetic field; however, this frequency range was too low to be picked up by the search coil.

**Results and Discussion**

Figure 3a presents the distribution of the peak frequencies recorded in the spacecraft frame and plotted against the angle, $\theta_{kV}$, between the inferred propagation direction and solar wind velocity. The solar wind velocity vector is defined as anti-parallel to the Sun-Saturn line. This is a good enough approximation since the angle of aberration at Saturn is only ~1°. The negative frequencies indicate left-handed polarization in the spacecraft frame. We observe 70% of the frequencies in the right-handed sense, having a median $\theta_{kV}$ of 71° (solid line) with 25th and 75th percentiles of 64° and 75° (dashed lines) respectively. The minority of frequencies in the left-handed sense have a median $\theta_{kV}$ of 58° with 25th and 75th percentiles of 38° and 65° respectively. This separation between handedness and size of $\theta_{kV}$ can be explained by the extent of the Doppler shift imposed on the whistler waves. The Doppler shift is maximum when the waves propagate exactly anti-parallel to the solar wind. With increasing $\theta_{kV}$, the action of the solar wind in reversing the handedness becomes less effective since the component of the solar wind vector anti-parallel to the whistler propagation direction decreases. The handedness is thus more likely to be observed in its intrinsic right-hand polarization. This is in contrast with what is seen in the near-Earth space where the majority of observations are left-handed [e.g. *Halekas et al.*, 2006; *Russell et al.*, 2007]. This was attributed to the typical $\theta_{kV}$ of ~45° at 1 AU.

In Figure 3b, we plot the distributions of two more angles, $\theta_{BV}$ and $\theta_{kB}$; the former defined as the angle between the magnetic field and solar wind velocity vector. The Parker spiral is corroborated by the near-perpendicular $\theta_{BV}$ with a median of 83°. The distribution of $\theta_{kB}$ indicates that whistler waves propagate close to the magnetic field direction with a median of 33°. This is supported by both theory and simulations which demonstrate that whistlers are able to propagate closer to the magnetic field [*Gurnett*, 1995].

*Russell* (2007) presented a set of power spectra revealing the peak frequencies of whistler waves upstream of each planet to decrease with heliocentric distance. This can be explained by the dispersion relation for whistler waves. For a cold proton-electron plasma, the phase and group speeds of whistlers derived from the dispersion relation are given by [*Stix*, 1962],

$$v_p \equiv \frac{\omega}{k} = \frac{c\sqrt{\omega\omega_{ce}}}{\omega_{pe}} \qquad (2a)$$

$$v_g \equiv \frac{\partial\omega}{\partial k} = 2v_p \qquad (2b)$$

Figure 4 presents this relationship graphically. The dispersion relations of the right-handed whistler branch are shown for typical conditions upstream of the Earth and Saturn. The number density and magnetic field used are 5 cm$^{-3}$ and 5 nT for the Earth and 0.5 cm$^{-3}$ and 1 nT for Saturn. The dispersion relations are in the form of phase and group speeds with respect to frequency, restricted to the ion and electron cyclotron frequencies for each planet. The lowest speeds are marked by the Alfvén speed with a lower speed at Saturn than at the Earth. This translates to a higher Alfvén Mach number at Saturn, assuming a constant solar wind speed. Overlaid on the phase and group speeds are the Doppler shift imposed by the solar wind for two cases $\theta_{kV} = 0°$ and 65°. The former is an end case where the Doppler shift effect is at a maximum. This is more likely at the Earth, though not necessarily the most typical (observationally a median of ~30° [*Russell.*, 2007]). The latter is what was inferred at Saturn. The combination of the upstream state of the solar wind and the geometry of the Parker spiral determines the peak frequency. The upstream density and magnetic field set the limits of the frequency range in which whistlers can propagate. This leads to a leftward and downward shift of the phase and group speeds from the Earth to Saturn.

As described above, whistlers propagating upstream must compete against the downstream-travelling solar wind. The group velocity at which whistlers can propagate upstream is marked by the interception between the solar wind speed and group speed in Figure 4. For each planet, whistlers are able to propagate upstream for any frequency rightward of this point (i.e. $v_g > \boldsymbol{V_{SW}} \cdot \boldsymbol{\hat{k}}$), as labelled by the solid arrows. Furthermore, the polarization depends on the interception between the solar wind speed and phase speed. The polarization is reversed leftward of this point, but always to the right of $v_g$, such that the information reaches the spacecraft but the phases are swept away downstream. Conversely, the polarization is unchanged rightward of this point where the phases are able to overcome the solar wind's Doppler shift. The median peak frequency for the subset of right-handed observations was higher at 0.35 Hz compared to 0.28 Hz for the left-handed. This is consistent with Figure 4 where the range of frequencies above $v_p$ are higher than the range of frequencies between $v_g$ and $v_p$. Figure 4 also seeks to explain the lower peak frequency at Saturn which has a typical value between 0.3 and 0.4 Hz in the spacecraft frame, in contrast with 1 Hz at Earth. With increasing heliocentric distance, whistler waves are able to radiate upstream forming a precursor wave train at a lower group speed since $v_g \propto \sqrt{\omega}$. This frequency represents the shortest wavelength capable of standing in the flow. We note that while the dispersion relation is for a cold plasma, i.e. more applicable at 10 AU where the species temperatures are significantly lower, it is strictly in the plasma frame. Thus, we use the dispersion relation to qualitatively explain the difference in frequencies.

In the context of dissipation, Landau damping is understood to be responsible. Whistlers are capable of interacting with electrons via trapping, which consequently leads to electron heating [*Burgess and Scholer*, 2015]. Simulations reveal $\beta_e$ (ratio of electron thermal to magnetic pressures) to be a control parameter with higher values making Landau damping more important [*Liewer et al.*, 1991]. Unlike the Mach number, $\beta$ increases only marginally between 1 and 10 AU, hence we do not expect Landau damping to play a significantly larger role at Saturn's bow shock. The propagation angle to the magnetic field is also a factor with stronger damping at greater $\theta_{kB}$ according to Landau resonance theory. The observations suggest that the small $\theta_{kB}$ measured is likely to allow the whistler waves to travel an appreciable distance upstream before being Landau damped. Despite these consistencies, we note that they are not exclusive to the interpretation of Landau damping. Other contributors to dissipation have been proposed such as amplification of the waves by large pitch-angle backstreaming electrons and the influence of scattering by ion acoustic turbulence [*Burgess and Scholer*, 2015].

The shock conservation relations dictate that the higher the Mach number, the more dissipation must take place within (and near) a collisionless shock in order to balance the energies upstream and downstream. The means by which the shock achieves this, as well as the partitioning of dissipative energy among different species, is beyond this classical framework. It has been well established that particle dynamics play a crucial role in achieving this dissipation [*Paschmann et al.*, 1981]. A fraction of incoming ions is decelerated and reflected back upstream by the shock's potential barrier. These reflected ions return to the shock after a partial gyration and are eventually transmitted with a strong perpendicular component of their velocity thus increasing the kinetic temperature of their distribution. Ion reflection almost always takes place at Saturn's bow shock, and recent observations have indicated that shock reformation occurs as a result [*Sulaiman et al.*, 2015]. Observations have also shown that although the total heating increases with Mach number, the proportion of the total energy used in heating the electrons falls [*Schwartz et al.* (1988); *Masters et al.* (2011)]. The complete picture of energy partitioning between the species remains an open question. Whistler waves are known to play a role in the electron dynamics, though their contribution to the relative heating of elections especially in this parameter space is beyond the scope of this work. We note that the low occurrence of whistler waves observed upstream may be a lower limit since their detection upstream is limited by several factors discussed earlier.

Observations first made at Earth and more recently extended to Saturn, have shown a relationship between $M_w$ and electron acceleration [*Oka et al.*, 2011; *Masters et al.*, 2016]. The spectral index of the electron energy spectrum was found to be controlled by $M_A/M_w$ - i.e. electron acceleration was prominent at supercritical shocks. This was interpreted as the electrons being accelerated at the shock instead of escaping upstream with the propagation of whistlers as they would be permitted in the subcritical case.

**Conclusion**

We have conducted a study characterizing upstream whistler-mode waves for a unique parameter space at 10 AU, in comparison with the near-Earth space. The results near Saturn show these waves are observed at much lower frequencies, between 0.2 and 0.4 Hz in the spacecraft frame, compared to the typical 1 Hz frequency near Earth. This is due to the lower ion and electron cyclotron frequencies near Saturn, between which whistler waves can develop. The polarization is also mostly observed in its intrinsic right-hand state, owing to the Parker spiral. The large angle between the solar wind speed and wave vector means the Doppler shift is significantly less and typically not sufficient to reverse the polarization in the spacecraft frame. This is the opposite case to the Earth where most are observed as left-handed in the spacecraft frame. Whistler waves contribute to the electron dynamics in shocks, however the total picture of heating and acceleration remains unclear particularly in the context of the lower proportion of the total dissipation in electron heating [*Schwartz et al.* (1988)]. For completeness, there may be scope, particularly in the Magnetospheric Multiscale (MMS) mission to address some of the open questions concerning the generation and role of whistlers in dispersive shocks.

We acknowledge the support of MAG data processing/distribution staff. Cassini magnetometer and RPWS data are publicly available via NASA's Planetary Data System. The research at the University of Iowa was supported by NASA through Contract 1415150 with the Jet Propulsion Laboratory.

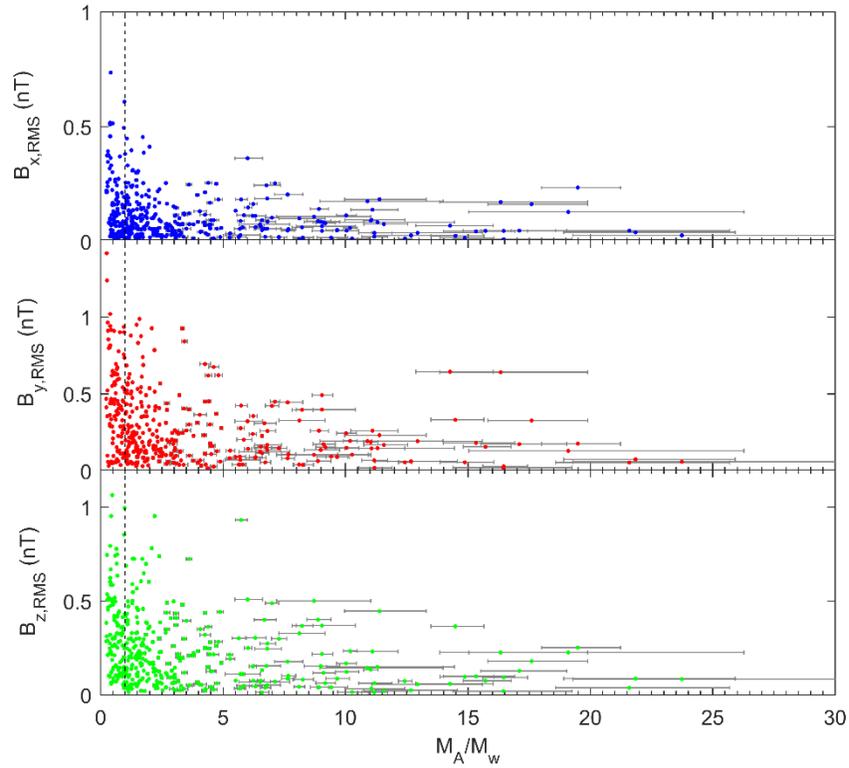

Figure 1: RMS of the components of the upstream magnetic field versus $M_A/M_W$ where $M_W$ is the critical Mach number above which whistler waves cannot propagate upstream. Dashed line marks the critical $M_A/M_W = 1$ separating the subcritical ($M_A < M_w$) from the supercritical ($M_A < M_w$) classes.

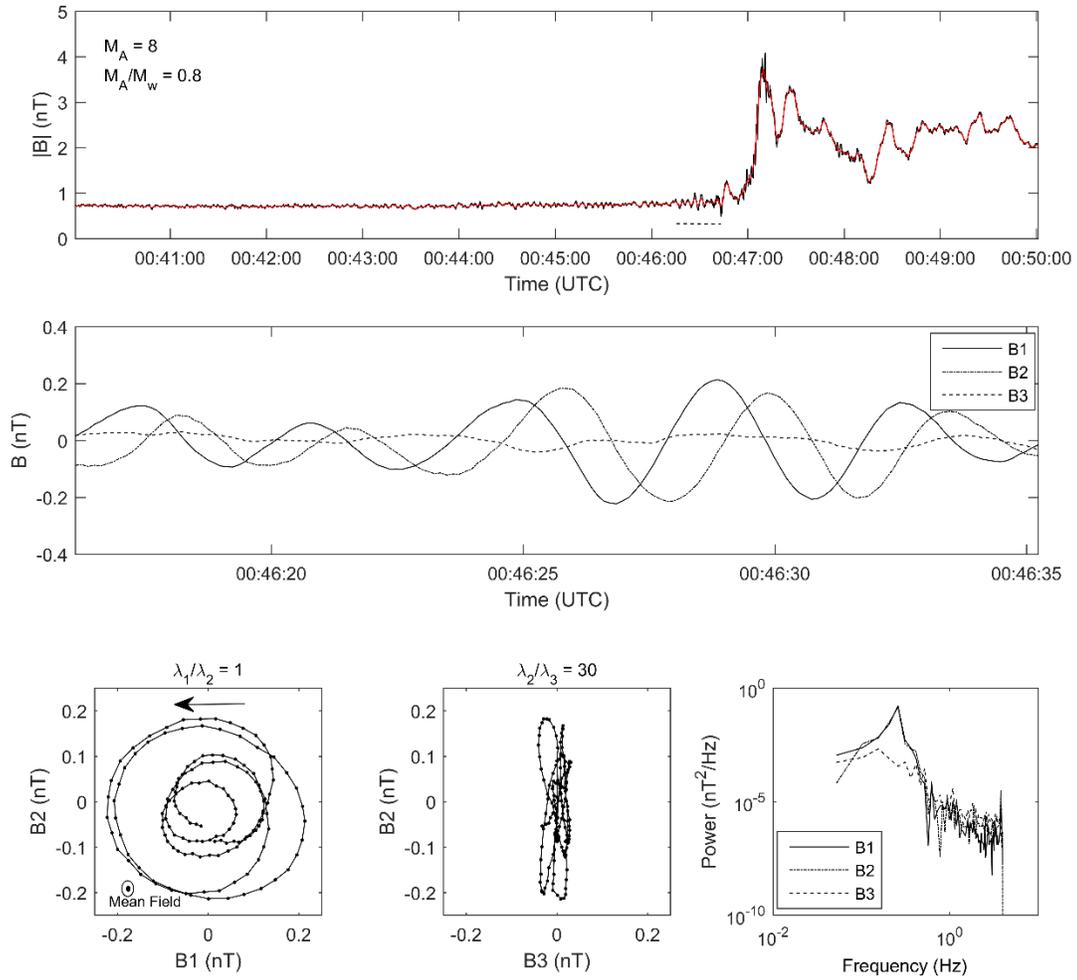

Figure 2 - Top: Magnetic field (black) and averaged (red) time-series of a subcritical quasi-perpendicular Saturnian shock. Middle: Upstream interval, underlined by dashed line, transformed into directions of maximum (B1), intermediate (B2), and minimum (B3) variances. Bottom: Hodograms of the interval and power spectrum.

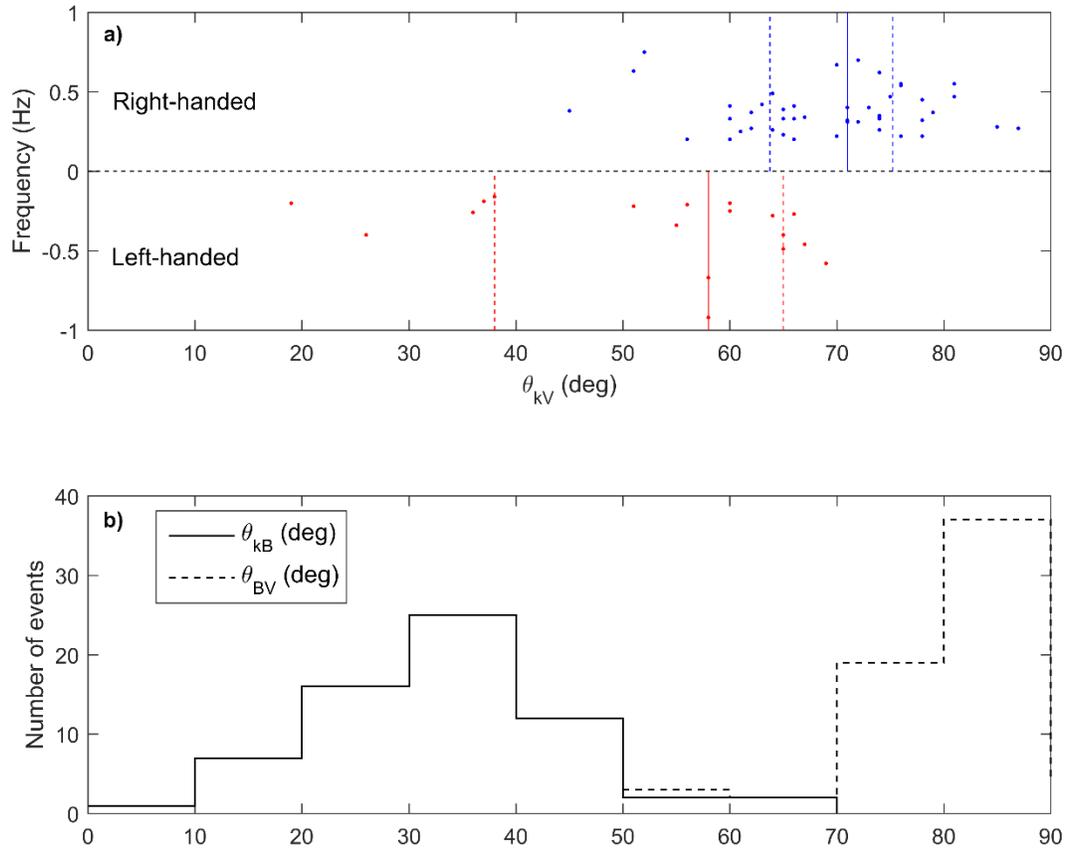

Figure 3 – a) Distribution of frequencies, separated by handedness, against $\theta_{kV.}$ (there is a sign ambiguity so all angles are referenced to the range 0-90°). b) Distribution of $\theta_{kB}$ and $\theta_{BV.}$

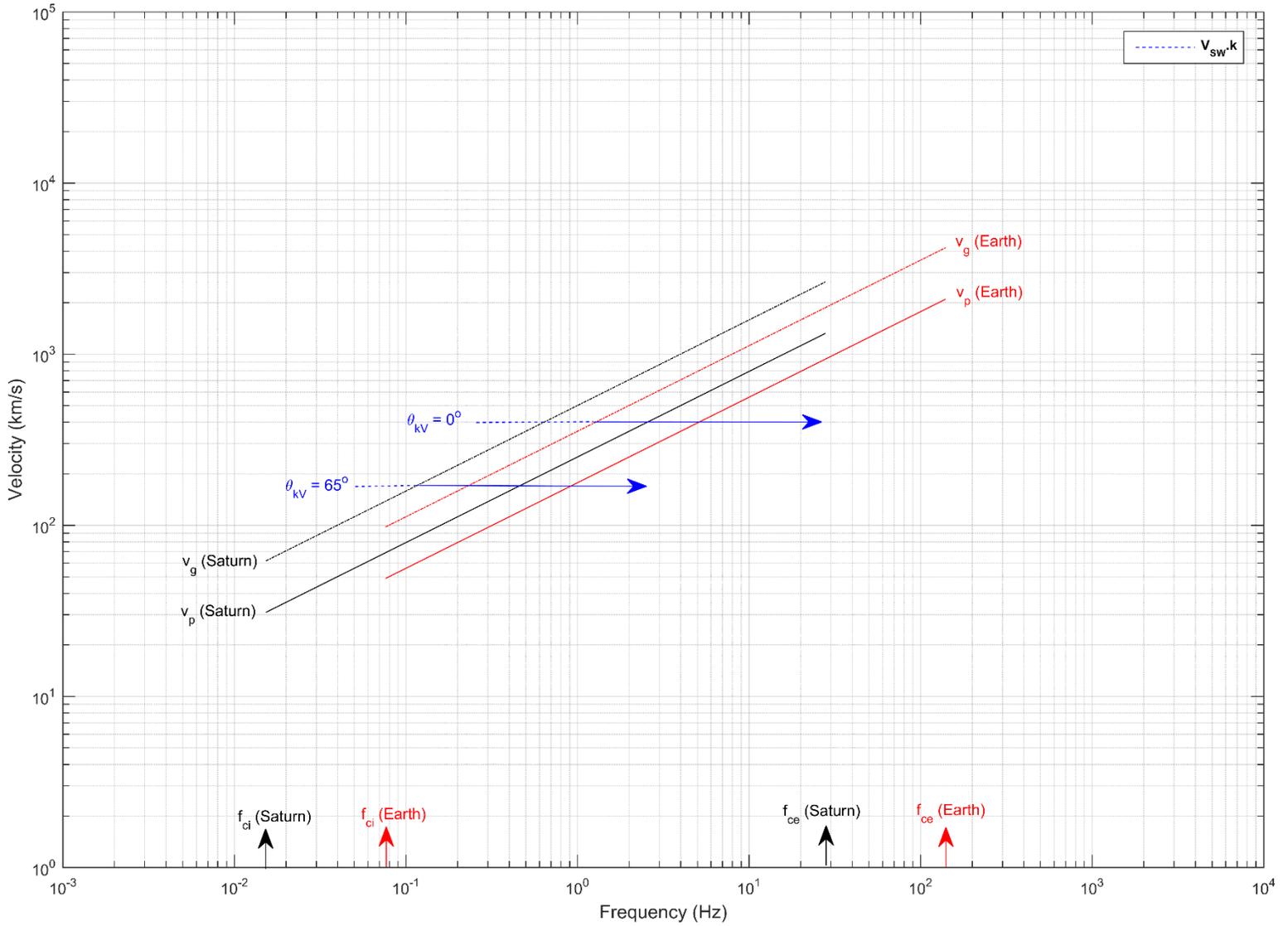

Figure 4 – Dispersion relation for whistlers in a proton-electron cold plasma with typical conditions at Earth (red) and Saturn (black). Group, $v_g$, and phase, $v_p$, velocities are plotted as a function of frequency for each planet between their respective ion and electron cyclotron frequencies. The blue plot is the Doppler shift imposed by the solar wind for different $\theta_{kV}$. The solid blue arrow represents the range of frequencies over which whistlers can propagate upstream for the given $\theta_{kV}$.